# Monolithic Integration of a Quantum Emitter with a Compact On-chip Beam-splitter

**N. Prtljaga[1,a], R. J. Coles[1], J. O'Hara[1], B. Royall[1], E. Clarke[2], A. M. Fox[1], and M. S. Skolnick[1]**

[1]*Department of Physics and Astronomy, University of Sheffield, S3 7RH, United Kingdom*
[2]*Department of Electronic and Electrical Engineering, University of Sheffield, Sheffield S1 3JD, United Kingdom*

A fundamental component of an integrated quantum optical circuit is an on-chip beam-splitter operating at the single-photon level. Here we demonstrate the monolithic integration of an on-demand quantum emitter in the form of a single self-assembled InGaAs quantum dot (QD) with a compact (>10 µm), air clad, free standing directional coupler acting as a beam-splitter for anti-bunched light. The device was tested by using single photons emitted by a QD embedded in one of the input arms of the device. We verified the single-photon nature of the QD signal by performing Hanbury Brown- Twiss (HBT) measurements and demonstrated single-photon beam splitting by cross-correlating the signal from the separate output ports of the directional coupler.

The proposal of linear optics quantum computing (LOQC) represents a significant step towards scalable optical quantum information processing[1]. Integrated optics offers a route forward to achieving the high component density that LOQC requires[2], while also improving the intrinsic stability of the circuitry and giving a significant reduction in the size and complexity of the experimental apparatus[3]. III-V semiconductor circuits offer additional benefits over other technologies in this context by providing a platform for incorporating on-demand quantum emitters such as self-assembled quantum dots (QDs) within the circuit, and also allowing for hybrid approaches beyond LOQC for example by forming spin-photon networks[2,4,5].

A fundamental component of an integrated quantum optical circuit is an on-chip beam-splitter operating at the single-photon level[1]. In this Letter we demonstrate the monolithic integration of an on-demand quantum emitter in the form of a single self-assembled InGaAs quantum dot with an on-chip directional coupler acting as a beam-splitter for anti-bunched light. We first discuss the design of the devices, and then present results demonstrating their performance at the single photon level.

The overall structure of the device is shown in Fig. 1(a). Directional couplers were used as the beam-splitter on account of their easy modeling, flexibility in the splitting ratio and low back-scattering losses[6]. In addition, they do not require tapering to interface them with waveguides or with more complex structures such as nanobeam cavities[7] or waveguide based spin-photon interfaces[8], which allows for low overall losses and relatively simple circuit designs.

---

[a] Author to whom correspondence should be addressed. Electronic mail: n.prtljaga@sheffield.ac.uk

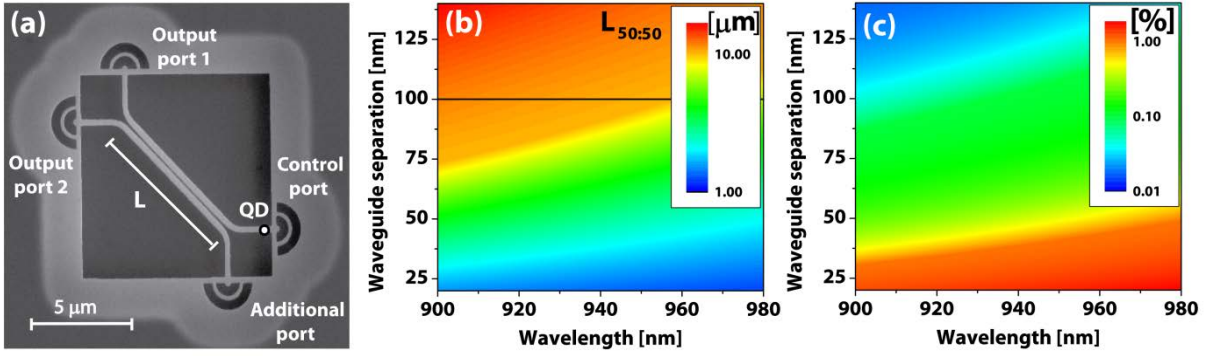

**FIG. 1.** (Color online) (a) SEM image of directional coupler. *L* is the interaction length between the waveguides. (b) Simulated logarithmic color map of 50:50 coupling length ($L_{50:50}$) as a function of wavelength and waveguide separation (waveguide width of 280 nm was used). Black line indicates waveguide separation of 100nm. (c) Simulated logarithmic color map of optical loss induced by mismatch (spatial and phase) between the input waveguide and the coupling region of the directional coupler. Optical loss is plotted as a function of wavelength and waveguide separation (waveguide width of 280 nm again used) and is normalized with respect to the input power.

The waveguide dimensions were chosen in order to assure operation in TE polarization. Optical simulations of the devices were performed using a commercial-grade eigenmode solver.

For weak coupling, the modes formed in the coupling region can be approximated as symmetric and anti-symmetric superpositions of individual waveguide eigenmodes commonly known as super-modes[9]. Coupled mode theory can then be used to obtain a simple analytical expression for the power transfer between the waveguides[9,10].

In the case of a directional coupler composed of two identical waveguides, the interaction length needed for 50:50 power transfer is given by[9–11]:

$$L_{50:50} = \frac{\lambda_0}{\pi * \Delta n} * \sin^{-1}\left(\sqrt{0.5}\right) \quad (1)$$

where $L_{50:50}$ is the interaction length at which 50% of the incident power is transmitted to the second waveguide, $\lambda_0$ is the free space wavelength and $\Delta n$ is the effective index difference between the symmetric and anti-symmetric super-modes and is calculated numerically using the eigenmode solver for a range of parameters of interest. Calculated values for $L_{50:50}$ are shown in Fig. 1(b). For a given waveguide separation and waveguide width, $L_{50:50}$ increases with decreasing wavelength (due to better mode confinement), indicating weaker coupling for shorter wavelengths. For waveguide separations smaller than 100 nm, $L_{50:50}$ is shorter than 10 µm. This allows for devices of a very small footprint of order 10-20 µm, comparable to recently demonstrated plasmonic circuits[12]. Similar footprint devices can also be realized in silicon-on-insulator (SOI)[13,14].

The main added losses in directional couplers compared to single waveguides of comparable length originate from the bends[6] together with spatial and phase mismatch between the input waveguides and the coupling region[16].

The strong mode confinement for high index contrast air clad GaAs waveguides allows very sharp waveguide bends. We find that the bending loss becomes negligible (< 1%) for bend radii of 2 µm or larger, a value which we use in the fabrication of our devices.

An estimate of the mode–conversion loss between the input waveguide and the coupling region can be obtained by calculating the spatial profile mismatch and Fresnel reflections due to the difference in propagation constants[11]. The results of these calculations for a wide range of parameters of interest are reported in Fig 1(c). The maximum loss reaches a value of a few percent only for a very small waveguide separation (~20 nm).

The samples were grown by molecular beam epitaxy (MBE) on undoped GaAs (100) wafer, comprising a 140nm GaAs layer containing a layer of self-assembled InGaAs QDs at its center, above a 1μm thick sacrificial $Al_{0.6}Ga_{0.4}As$ layer on an undoped GaAs substrate. The devices were fabricated by electron beam lithography, followed by an inductively-coupled plasma etch to define the pattern into the GaAs membrane. The $Al_{0.6}Ga_{0.4}As$ layer was removed by an isotropic hydrofluoric acid etch to leave freestanding air-clad GaAs waveguides. A scanning electron microscope (SEM) image of the fabricated device is shown in Fig. 1(a). Semicircular air/GaAs grating output couplers[16] were added to the end of the waveguides to scatter light out of the device plane into the detection apparatus.

Optical measurements were performed at 4.2 K in a confocal microscope system with three independent optical paths (one excitation and two collection paths). The photoluminescence (PL) was excited using an 850 nm continuous wave (CW) laser focused to a spot of diameter of ~1 μm. The PL was collected from the output couplers and filtered independently in each collection path both spectrally and spatially[17]. For detection single-photon avalanche photodiodes (SPADs) or a charge coupled device (CCD) camera were used.

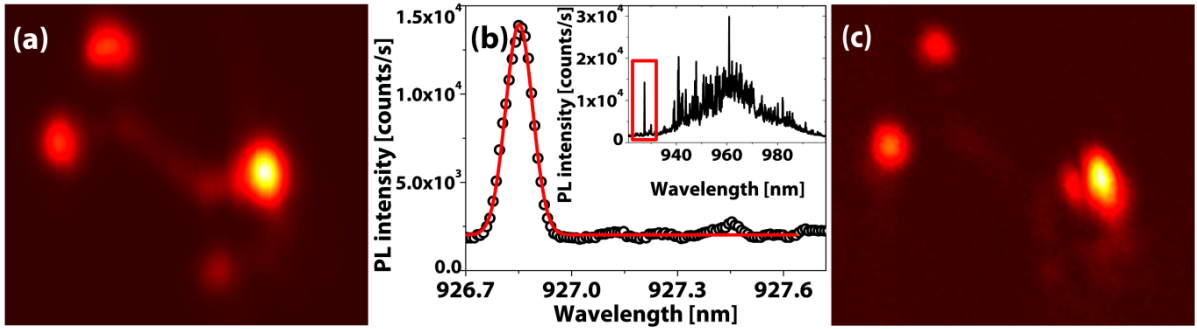

**FIG. 2 (Color online) (a) Ensemble PL map of investigated device. (b) Spectrum of the QD that was used to characterize the directional coupler at the single photon level. Fit to the experimental data yields a spectrometer limited linewidth of 0.09 nm (~130 μeV). Inset: Spectrum of ensemble PL as seen from control port for excitation spot position which yields maximum intensity for QD line indicated by red rectangle. (c) PL map of the same device as in panel (a) obtained by spectrally filtering the collected light (0.17 nm spectral bandwidth) at the peak of QD emission indicated in panel (b) of this figure.**

The device was initially characterized using the QD ensemble PL acting as an internal light source. The excitation spot was positioned at the waveguide end close to the control port (see Fig. 1(a)). Fig. 2(a) shows the PL map obtained by raster scanning the collection spot for a device with $L$=7 μm. It is apparent that the light generated at the control port is coupled to one arm of the directional coupler and then equally divided between the output ports (Fig. 2(a)), demonstrating 50:50 splitter operation for the ensemble PL

The device was then tested at the single-photon level by working with a spectrally isolated QD emitting at 927 nm on the blue side of the QD distribution (Fig 2(b)). The QD was located near the control port, as indicated in Fig. 1(a). Modifications to the local density of optical states (LDOS) in the high index-contrast structure lead to efficient funneling of QD photons into the waveguide. By performing 3D Finite-difference time-domain (FDTD) simulations, the coupling efficiency of the QD emission to the waveguide propagating mode was deduced. For a QD positioned at the exact centre of the waveguide, a coupling efficiency of ~95 % is found. The coupling efficiency remains higher than >80% even for significant (up to 100 nm) QD displacement from the waveguide centre. The remaining fraction is radiated perpendicular to the waveguide direction (horizontally and vertically).

Photons that are coupled to the waveguide mode travel either forwards or backwards with equal probabilities. Those travelling towards the control port were used to monitor the single-photon nature of the QD emission, while the others were used to test the performance of the directional coupler.

The results in Fig 2(a) show that the device acted as a 50:50 beam-splitter around the peak of the ensemble PL (960 nm), but this did not hold for the wavelength of the dot (927 nm) where the splitting ratio was close to 70:30 (Fig. 2(c)). Generally, the coupling between the waveguides decreases for shorter wavelengths due to the better mode confinement leading to longer $L_{50:50}$ (see Fig. 1(b)). The fact that we observe stronger coupling for shorter wavelengths indicates that this particular device was operating beyond its first 50:50 coupling point. In an HBT experiment, an unbalanced beam-splitter can always be converted to a balanced one by introducing additional optical loss to one of the optical paths before the detectors. This was the strategy adopted here; the effect of imbalance manifests itself only in a reduced number of coincidence counts.

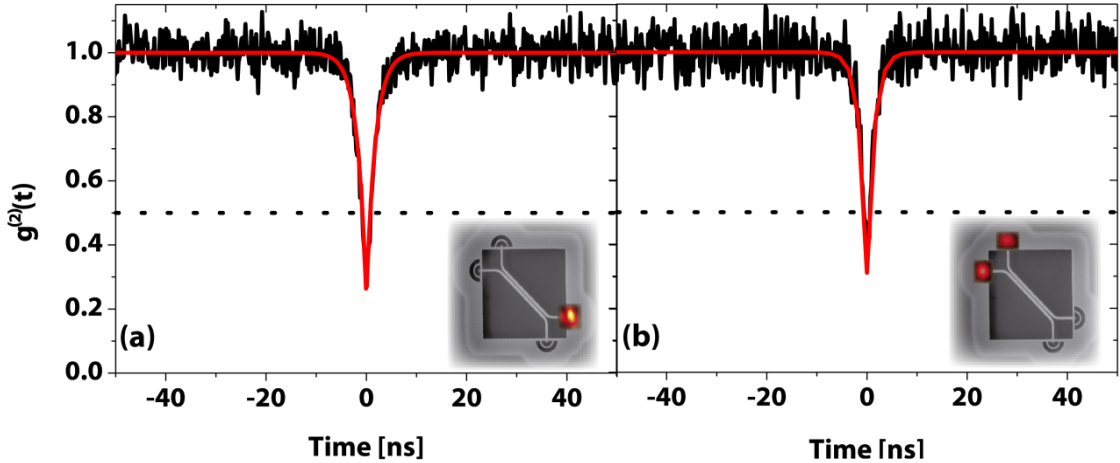

**FIG. 3 (a) Normalised auto-correlation function of QD signal (black line, without background subtraction) taken from control port (inset). Dotted line indicates 0.5 limit. Red line is the best fit to the experimental data yielding $g^{(2)}(0) = 0.23\pm0.02$. (b) Normalised cross-correlation function (black line, without background subtraction) from two different output ports (inset). Dotted line indicates 0.5 limit. Red line is the best fit to experimental data yielding $g^{(2)}(0) = 0.31\pm0.03$.**

The result of an HBT measurement performed directly on the dot via photons emitted from the control port is shown in Fig. 3(a). A fit to the normalized data without background subtraction gives $g^{(2)}(0) = 0.23\pm0.02$ (Fig. 3(a)). $g^{(2)}(0)$ does not reach zero due to the background from the other QDs and the limited timing resolution of our system (~520 ps). This can be compared to the results in Fig. 3(b), which show the cross-correlated QD signal from the output ports. In this case, the directional coupler acts as the beam splitter of the HBT experiment. A value of $g^{(2)}(0) = 0.31\pm0.03$ is found, again below 0.5, demonstrating on-chip HBT functionality at the single-photon level. When corrected for the background and temporal response of the system, a value of $g^{(2)}(0) = 0.10\pm0.05$ is found.

In conclusion, we have demonstrated the monolithic integration of an on-demand quantum emitter in the form of a single self-assembled InGaAs quantum dot (QD) with an air clad, free standing directional coupler. By careful optimization of the air gap and the length of the coupling region, we realized an on–chip 50:50 beam splitter. We tested the device at the single photon level by using the anti-bunched light from a single QD embedded in the waveguide and demonstrate single-photon beam splitting by cross-correlating the QD signal from separate output ports. This work paves the way towards demonstration of on-chip quantum optical circuits with monolithically integrated on-demand quantum emitters.

The authors would like to thank I. J. Luxmoore for fruitful discussion. This work was funded by EPSRC grant number: EP/J007544/1.